\documentclass[twocolumn,journal]{IEEEtran}
\usepackage[T1]{fontenc}
\usepackage[latin9]{inputenc}
\usepackage{array}
\usepackage{calc}
\usepackage{multirow}
\usepackage{amsmath}
\usepackage{graphicx}
\usepackage[unicode=true,
 bookmarks=true,bookmarksnumbered=true,bookmarksopen=true,bookmarksopenlevel=1,
 breaklinks=false,pdfborder={0 0 0},pdfborderstyle={},backref=false,colorlinks=false]
 {hyperref}
\hypersetup{pdftitle={Your Title},
 pdfauthor={Your Name},
 pdfpagelayout=OneColumn, pdfnewwindow=true, pdfstartview=XYZ, plainpages=false}
\usepackage{breakurl}
\makeatletter

\providecommand{\tabularnewline}{\\}

 \let\oldforeign@language\foreign@language
 \DeclareRobustCommand{\foreign@language}[1]{%
   \lowercase{\oldforeign@language{#1}}}

\usepackage[caption=false,font=footnotesize]{subfig}

\makeatother

\begin{document}

\title{Noise Attention based Spectrum Anomaly Detection Method for Unauthorized
Bands}

\author{Jing Xu, Yu Tian, Shuai Yuan and Naijin Liu \thanks{Jing Xu, Yu Tian, Shuai Yuan and Naijin Liu are with Qian Xuesen Laboratory of Space Technology, China Academy of Space Technology, Beijing 100094, China, e-mail: \protect\href{http://xxx@xxx.xxx}{xj\_nwpu@163.com}; \protect\href{http://xxx@xxx.xxx}{tianyu@qxslab.cn}; \protect\href{http://xxx@xxx.xxx}{yuanshuai@qxslab.cn}; \protect\href{http://xxx@xxx.xxx}{liunaijin@qxslab.cn}. (Corresponding author: Yu Tian.)}}


\maketitle
\begin{abstract}
Spectrum anomaly detection is of great importance in wireless communication
to secure safety and improve spectrum efficiency. However, spectrum
anomaly detection faces many difficulties, especially in unauthorized
frequency bands. For example, the composition of unauthorized frequency
bands is very complex and the abnormal usage patterns are unknown
in prior. In this paper, a noise attention based method is proposed
for unsupervised spectrum anomaly detection in unauthorized bands.
First, we theoretically prove that the anomalies in unauthorized bands
will raise the noise floor of spectrogram after VAE reconstruction.
Then, we introduce a novel anomaly metric named as noise attention
score to more effectively capture spectrum anomaly. The effectiveness
of the proposed method is experimentally verified in 2.4 GHz ISM band.
Leveraging the noise attention score, the AUC metric of anomaly detection
is increased by 0.193. The proposed method is beneficial to reliably
detecting abnormal spectrum while keeping low false alarm rate.
\end{abstract}


\section{Introduction\label{sec:Introduction}}

\IEEEPARstart{W}{ith} the rapid development of radio technology,
many fields rely on spectrum to realize their function, resulting
in a greatly enlarged demand for radio spectrum resources. Besides,
spectrum is an open environment, where equipment can easily access,
which may expose the spectrum to the risks of various attacks and
severely affect the normal use of spectrum. For example, unintentional
and intentional interference to positioning service such as Global
Navigation Satellite System (GNSS), which is widely used in applications
like automatic vehicle navigation, aircraft landing and marine vessel
tracking, has increased due to the accessibility of interference equipment
\cite{ThombreGNSS,Strohmeier}. Meanwhile, illegal repeaters used
to enhance mobile coverage may adversely affect the mobile operator's
cell planning, resulting in poor coverage and dropouts. Therefore,
it requires the detection of threatened signals to secure safety.
Spectrum anomaly detection is such an essential approach to secure
safety, which can detect threatened spectrum in time so that it can
be easily eliminated later.

The research of spectrum anomaly detection can be divided into detection
in authorized bands and unauthorized bands. In authorized bands, only
authorized systems can legally access to certain frequency bands.
Therefore, the signal wave should be detected as abnormal spectrum
if it is different from that authorized systems. Thus, spectrum anomaly
detection in authorized frequency can be well solved by signal identification,
which has been well studied in recent years \cite{Salcedo2010Signal,Ebrahimzadeh2007Digital,Gorcin2012Template}.
The real challenge remains in detecting anomaly in unauthorized frequency
bands. In unauthorized bands, spectrum is open for access, and systems
with different center frequency and MAC coexist in this band. As a
result, the time and frequency of signal appearance are random, and
there is no prior information for anomaly detection. Existing detection
methods cannot detect the anomalies in the open frequency band effectively.
To solve this problem, our research tries to achieve the detection
of anomalies in unauthorized bands.

Abnormal detection in unauthorized bands faces many problems. First,
the spectrum data is very complex and it is hard to identify all the
abnormal patterns, thus manual spectrum labeling is very difficult.
Second, since the number of the spectrum data is extremely large but
rare of them are abnormal, it is hard to collect sufficient abnormal
samples for training. Therefore, the application of supervised learning
methods in radio spectrum anomaly detection is highly limited due
to the loss of labeled data, and a complete unsupervised, automatic
feature-extracted learning model is highly required to solve this
problem.

In this paper, a noise attention based method is proposed to solve
the abnormal spectrum detection problem in wireless communication
environment. First, received signal is pre-processed to obtain its
spectrogram, which intuitively reflects the correlation of signals
over different time and frequency. Second, we only use normal spectrum
data to train the deep VAE so that the model will capture the characteristics
of normal spectrum. Finally, a novel anomaly score is designed to
better capture the physical nature of input signals and to effectively
detect abnormal signals. Experiment results show that the proposed
method could greatly improve the performance of anomaly detection
of spectrum in wireless communication.

The rest of this paper is organized as follows. In Sec. \ref{sec:Related-works},
the previous works in spectrum anomaly detection are briefly introduced.
Then, Sec. \ref{sec:Principle} studies the principle of the reconstructed
output and the reconstructed abnormal output in spectrum data set
obtained from VAE. Then, in Sec. \ref{sec:Experimental-results} experimental
tests are designed to verify the high-performance of the proposed
method in this section. Finally, the main conclusion of the research
is summarized in Sec. \ref{sec:Conclusions}.

\section{RelatedWork\label{sec:Related-works}}

Numerous researches have been conducted on spectrum anomaly detection,
which can be divided into two categories with respect of their pattern
extraction methods. The first kind uses manually patterns designed
by experts and the second uses pattern obtained through learning. 

Refs. \cite{2010Spectrum,Liu2009ALDO,LiuDetecting,Yinsixing} are
the methods that use manual patterns. Ref. \cite{2010Spectrum} used
database comparison method, which discovered anomaly by comparing
real signal with the standard wave patterns pre-existing in the database,
such as frequency, occupancy, field strength, bandwidth, direction,
polarization and modulation. However, it was difficult to construct
a complete database. In Ref. \cite{Liu2009ALDO}, the spectrum anomaly
detection problem was converted into a statistical significance testing
problem. It leveraged the property that the received signal power
(RSS) decays approximately linearly with the logarithmic distance
from the source and then unauthorized transmitters can be detected
by making use of the propagation characteristics. Ref. \cite{LiuDetecting}
also made use of the property that transmitters at different locations
will lead to different spatial distributions of the RSS and detect
anomalies by comparing the current pattern with a stored spatial map
of the transmitter. Ref. \cite{Yinsixing} obtained the historical
pattern by calculating the average of seven days' data. It calculated
Mahalanobis distance between measuring spectrum and the historical
pattern to detect potential anomalies.

On the other hand, considerable researches leverage data-driven learning
methods to select patterns. Ref. \cite{FengAnomaly,unsupervised}
used a deep auto-encoder model to perform normal pattern extraction.
In their paper, spectrogram was used as the input of the learning
model. The anomaly score it used was mean squared error (MSE) between
the amplitude across sub-frequencies of the true spectrogram and the
corresponding reconstructed one. To detect spectrum anomalies, Ref.
\cite{Wei2015Spectrum} adopted the idea of classification, using
spectrum occupancy sequence as input. Normal and abnormal spectrum
patterns were modeled by hidden Markov model (HMM). Therefore, the
anomalies can be recognized by computing maximum log-likelihood of
the data with respect to each spectrum pattern. Ref. \cite{ORecurrent}
used a long short-term memory (LSTM) based recurrent network to train
a time series model and to obtain features. Firstly, it computed the
difference between predicted value and true value on training set.
Then, it modeled the error vector using a parametric multivariate
Gaussian distribution. Finally, it computed the likelihood probability
in expected error distribution on test set to distinguish anomaly.
Ref. \cite{scaling} studied spectrum anomaly detection in LTE band.
It built deep neural network (DNN) models to capture spectrum usage
patterns and computed root mean squared error (RMSE) between the true
FFT amplitude across sub-frequencies and the model prediction values.

Most of the current researches on spectrum anomaly detection adopted
the idea of global average. In this paper, we propose a novel anomaly
score with its own attention mechanism, which assigns different weights
to different time-frequency regions when evaluating the degree of
abnormality. 

\section{Principle\label{sec:Principle}}

In this section, the training process of VAE\cite{Kingma} is briefly
described. Then, we reinterpret VAE training from the view of generation
error minimization and analyze the optimal output of VAE given certain
hidden variable. Finally, based on the property of VAE optimal output,
a novel anomaly score is designed for anomaly detection purpose. 

\begin{figure}[tbh]
\begin{centering}
\includegraphics[width=1\linewidth]{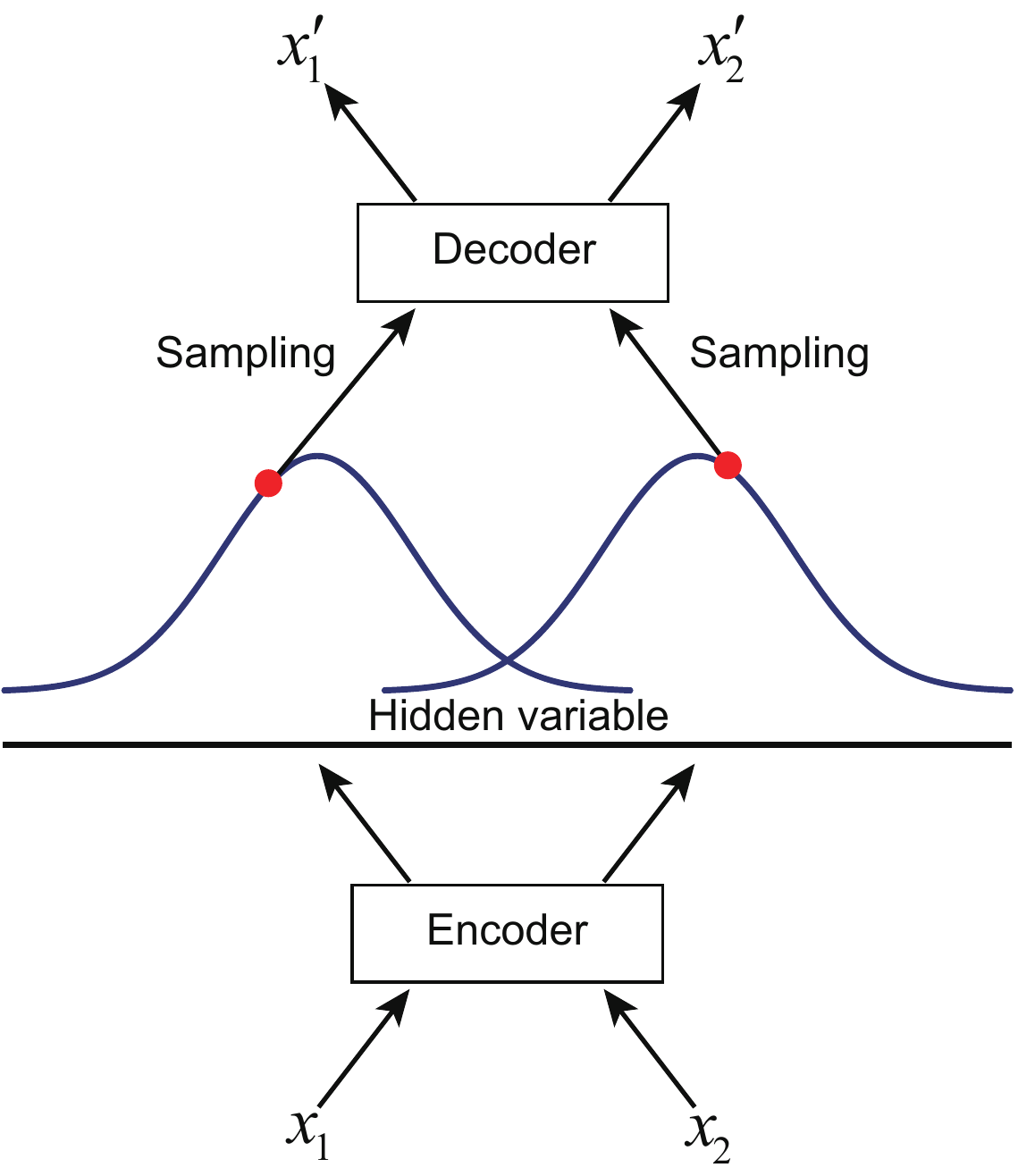}
\par\end{centering}
\caption{\label{fig:principle}The data flow of VAE.}
\end{figure}

As illustrated in Fig. \ref{fig:principle}, the training process
of VAE can be described as follows: the input $x^{i}$ is first passed
through the encoder to obtain the posterior probability $p_{\phi}\left(z|x^{i}\right)$,
then the hidden variable $z$ is sampled from the posterior probability,
after which the output $x_{i}^{\prime}=f_{\theta}\left(z\right)$
is obtained through the decoder. The objective function of VAE is
defined as follow, which equals to the marginal likelihood of $x_{i}$
estimated by the model.
\begin{align}
 & O\left(\phi,\,\theta;\,x_{i}\right)\nonumber \\
= & \frac{1}{2}\sum_{j=1}^{J}\left(1+\mathrm{log}\left(\left(\sigma_{i,j}\right)^{2}\right)-\left(\mu_{i,j}\right)^{2}-\left(\sigma_{i,j}\right)^{2}\right)\nonumber \\
 & +\sum_{l=1}^{L}p_{\phi}\left(z_{l}|x_{i}\right)\left[x_{i}-f_{\theta}\left(z_{l}\right)\right]^{2}\,,\label{eq:L}
\end{align}
where $\phi$ and $\theta$ are respectively the parameters of encoder
and decoder, $\sigma$ and $\mu$ represent the variance and mean
of hidden variable $z$, $i$ represents the $i$th sample of training
set, $L$ represent the number of sampling and l is the $l$th sampling.

The training process of VAE aims at minimizing the loss on training
set, which is the expectation of marginal likelihood over all training
data,
\begin{align}
 & \frac{1}{I}\sum_{i=1}^{I}O\left(\phi,\,\theta;\,x_{i}\right)\nonumber \\
= & \frac{1}{I}\sum_{i=1}^{I}\left\{ M_{i}+\frac{1}{I}\sum_{l=1}^{L}p_{\phi}\left(z_{l}|x_{i}\right)\left[x_{i}-f_{\theta}\left(z_{l}\right)\right]^{2}\right\} \nonumber \\
= & \frac{1}{I}\sum_{i=1}^{I}M_{i}+\frac{1}{I}\sum_{i=1}^{I}\sum_{l=1}^{L}p_{\phi}\left(z_{l}|x_{i}\right)\left[x_{i}-f_{\theta}\left(z_{l}\right)\right]^{2}\,,\label{eq:EL}
\end{align}
where $M_{i}=\frac{1}{2}\sum_{j=1}^{J}\left(1+\mathrm{log}\left(\left(\sigma_{i,j}\right)^{2}\right)-\left(\mu_{i,j}\right)^{2}-\left(\sigma_{i,j}\right)^{2}\right)$,
and $I$ is the number of samples in the training set.

In order to carry out our analysis, we assume $x_{i}$ is discrete.
The analysis result can be easily extended for continuous cases by
replacing the sum by integral. The number of $x_{i}$ with same value
$v_{n}$ is $c_{n}$,$n\in\left[1,\,N\right]$. Substituting $c_{n}$
into Eq. (\ref{eq:EL}),
\begin{align}
 & \frac{1}{I}\sum_{i=1}^{I}O\left(\phi,\,\theta;\,x_{i}\right)\nonumber \\
= & \frac{1}{I}\sum_{i=1}^{I}M_{i}+\sum_{n=1}^{N}\frac{c_{n}}{I}\sum_{l=1}^{L}p_{\phi}\left(z_{l}|v_{n}\right)\left[v_{n}-f_{\theta}\left(z_{l}\right)\right]^{2}\nonumber \\
= & \frac{1}{I}\sum_{i=1}^{I}M_{i}+\sum_{n=1}^{N}p_{\phi}\left(v_{n}\right)\sum_{l=1}^{L}p_{\phi}\left(z_{l}|v_{n}\right)\left[v_{n}-f_{\theta}\left(z_{l}\right)\right]^{2}\,,
\end{align}
where the second term of Eq. (\ref{eq:EL}) can be expanded as
\begin{align}
 & \sum_{n=1}^{N}p_{\phi}\left(v_{n}\right)\sum_{l=1}^{L}p_{\phi}\left(z_{l}|v_{n}\right)\left[v_{n}-f_{\theta}\left(z_{l}\right)\right]^{2}\nonumber \\
= & \sum_{n=1}^{N}\sum_{l=1}^{L}p_{\phi}\left(v_{n}\right)p_{\phi}\left(z_{l}|v_{n}\right)\left[v_{n}-f_{\theta}\left(z_{l}\right)\right]^{2}\nonumber \\
= & \sum_{n=1}^{N}\sum_{l=1}^{L}p_{\phi}\left(v_{n}|z_{l}\right)p_{\phi}\left(z_{l}\right)\left[v_{n}-f_{\theta}\left(z_{l}\right)\right]^{2}\nonumber \\
= & \sum_{l=1}^{L}p_{\phi}\left(z_{l}\right)\sum_{n=1}^{N}p_{\phi}\left(v_{n}|z_{l}\right)\left[v_{n}-f_{\theta}\left(z_{l}\right)\right]^{2}\,.
\end{align}
We define the generation error as
\begin{equation}
G(z_{l})=\sum_{n=1}^{N}p_{\phi}(v_{n}|z_{l})\left[v_{n}-f_{\theta}\left(z_{l}\right)\right]^{2}\,.\label{eq:generation err}
\end{equation}

For any encoder output $z_{l}$, there exist many different possible
input values $v_{n}$. Each $v_{n}$ will introduce a reconstruction
error $\left[v_{n}-f_{\theta}\left(z_{l}\right)\right]^{2}$, and
the generation error represents the probability weighting of all possible
reconstruction errors.

Substituting $G\left(z_{l}\right)$ into Eq. (\ref{eq:EL}), the loss
of VAE can be interpreted as
\begin{equation}
L\left(\phi,\,\theta\right)=\frac{1}{I}\sum_{i=1}^{I}M_{i}+\sum_{l=1}^{L}p_{\phi}\left(z_{l}\right)G(z_{l})\,,
\end{equation}
where $M_{i}$ represents the regularization term, which is used to
ensure an effective posterior probability for sampling $z_{l}$. After
training, the optimal decoder will minimize the generation error.
The property of optimal decoder is theoretically derived as follows.

As can be observed from Eq. (\ref{eq:generation err}), generation
error $G(z_{l})$ is actually a function of the decoder $f_{\theta}\left(z_{l}\right)$.
Given $z_{l}$, the optimal decoder gives the minimal generation error.
In order to obtain the extreme point, we differentiate $G(z_{l})$,
\begin{equation}
\frac{\partial G\left(z_{l}\right)}{\partial f_{\theta}\left(z_{l}\right)}=2\sum_{n=1}^{N}p_{\phi}(v_{n}|z_{l})\left[f_{\theta}\left(z_{l}\right)-v_{n}\right]=0\,.
\end{equation}
The optimal decoder $f_{\theta}\left(z_{l}\right)$ can be obtained
by letting above derivative equal to 0,
\begin{equation}
f_{\theta}^{*}\left(z_{l}\right)=\frac{\sum_{n=1}^{N}p_{\phi}(v_{n}|z_{l})v_{n}}{\sum_{n=1}^{N}p_{\phi}(v_{n}|z_{l})}\,\text{.}\label{eq:weighted sum}
\end{equation}
where it can be clearly observed that the trained decoder is approximately
the weighted average of all possible inputs. 

For abnormal samples, the hidden variable $z$ should be located in
the gap of the hidden variables that belongs to normal samples. The
reason behinds this is intuitive: if $z$ is not in the gap of the
normal hidden variables, then it must be close to one of hidden variables
of the normal sample. This means that this sample is very similar
to certain normal sample and it is not likely to be abnormal. 

When z locates in the gap, $p_{\phi}(v_{n}|z_{l})\ll1$, which means
none of the normal value $v_{n}$ can dominate the weighted sum defined
in Eq. (\ref{eq:weighted sum}). As a result, the decoder output approximates
the mean-average of nearby normal samples. In unauthorized frequency
bands, the access systems are various, and signals appears at random
time and frequency. For certain time-frequency area, the amplitude
behaves randomly over different samples, and the weighted sum defined
in Eq. (\ref{eq:weighted sum}) gives a mean amplitude. As a result,
the noise floor of abnormal samples is increased after reconstructed
by VAE model. Instead, when z belongs to normal samples, it will locate
within the realm of a normal sample. In this case, the probability
weight of certain sample should be very large, decoder output approximates
this normal sample.

On this basis, noise attention score is proposed to detect the abnormal
spectrum usage in unauthorized frequency bands,
\begin{equation}
N=\sum_{i=1}^{N}\frac{\left|x_{i}-\hat{x_{i}}\right|}{x_{i}}\,,
\end{equation}
where $x$ is the input amplitude of the variation encoder, $\hat{x}$
is the output amplitude of the variation encoder and $N$ is number
of pixels in the spectrogram. Compared to the traditionally used reconstruction
error score,
\begin{equation}
R=\sum_{i=1}^{N}\left|x_{i}-\hat{x_{i}}\right|\,,
\end{equation}
noise attention score pays more attention to noise area (low amplitude),
and is more sensitive to the noise floor change caused by abnormal
spectrum usage. For this reason, noise attention score can detect
abnormal spectrum more effectively and is used in our methodology.

\section{Experimental results\label{sec:Experimental-results}}

\subsection{Dataset\label{subsec:Dataset}}

In this paper, ISM band is chosen as the unauthorized band for spectrum
anomaly detection. The center frequency is set to be 2.4GHz and the
bandwidth is set to be 25MHz. WiFi, Bluetooth and etc. are included
in this observation. The USRP device is configured at a sampling rate
of 50 MSamp/s with omni-directional antennas which are deployed indoors.
The collected I/Q data was processed with 1024 points Short-Time Fourier
Transform (STFT), and then was down-sampled with a factor of 4. The
window of STFT is $1024$, the overlap is 0 and the sampling frequency
is $50\mathrm{MHz}$. The size of the final spectrogram images is
64 x 64. The frequency resolution is 200 kHz, the time resolution
is 80us, and the time span is 20ms.

\begin{figure}[tbh]
\begin{centering}
\includegraphics[width=1\linewidth]{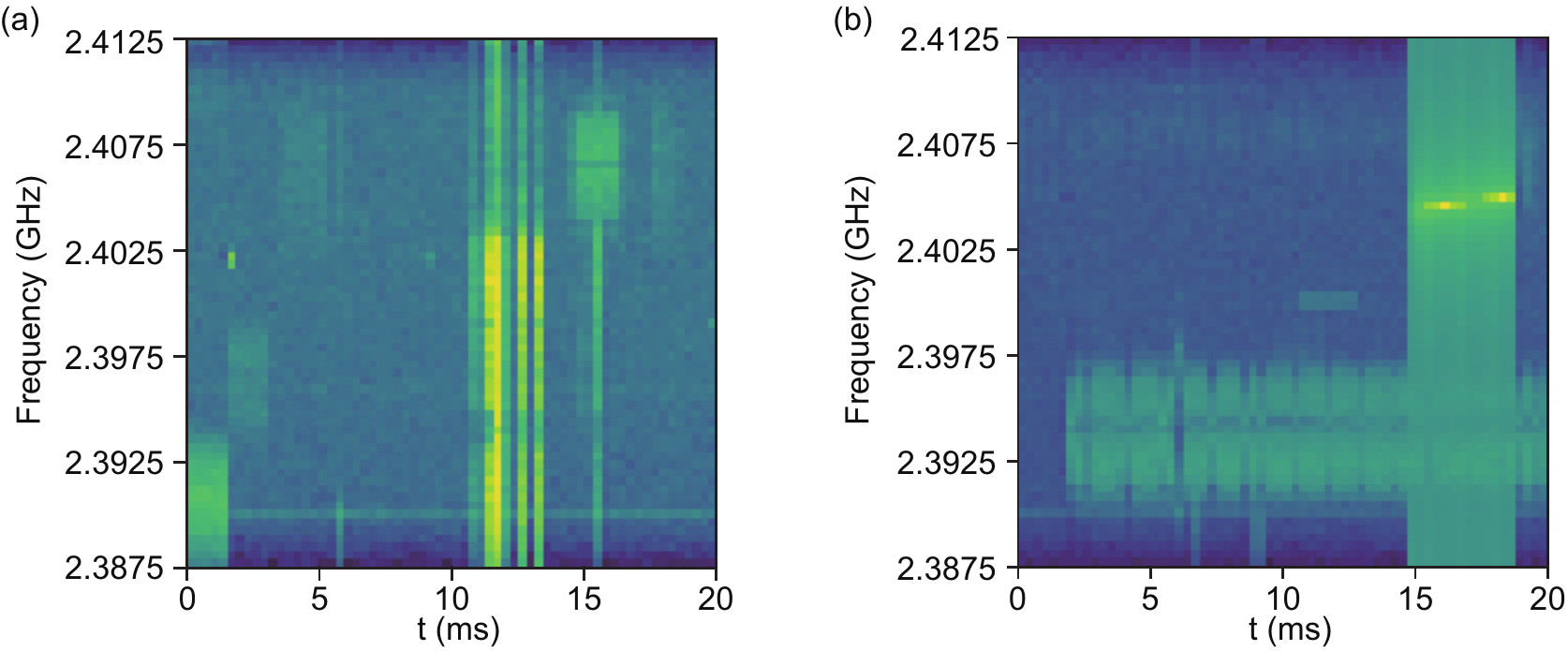}
\par\end{centering}
\caption{\label{fig:Spectrograms}Spectrograms of (a) normal and (b) abnormal
signals.}
\end{figure}

Spectrogram data is collected for the analysis of the proposed framework
over the period from 14:41:00 December 21, 2018 to 15:22:00 December
21, 2018. The collected spectrogram images are normal samples and
are randomly split into training set and test set, which contains
8438 and 906 spectrogram images respectively. In order to evaluate
the anomaly detection performance, anomaly samples have to be obtained
by adding anomaly signal into the normal samples. Ref. \cite{FengAnomaly}
adopted additive white Gaussian noise as anomaly signal to simulate
the anomaly case of the sudden signal-to-noise ratio change of the
communication channel. Refs. \cite{unsupervised} and \cite{2019Crowdsourced}
consider random bandwidth and SNR with deterministic shifts/hops in
frequency as anomaly. In this paper , we follow Ref. \cite{ORecurrent}
and define chirp signal as anomaly signal. Therefore, the test set
is further equally divided into two parts. The first part consists
of abnormal samples obtained by adding random chirp signals during
normal use \cite{ORecurrent}, while the second part remain unchanged
and is labeled as normal spectrograms. Examples of the normal and
abnormal spectrogram are shown in Fig. \ref{fig:Spectrograms}(a)
and (b).

With the obtained dataset, the variation encoder network is trained
using the train samples that are normal samples. After that, the test
samples consisting of normal and abnormal signals are fed into the
trained model to obtain the anomaly score. Finally, the performance
of the model can be verified by Receiver Operating Characteristic
(ROC) method. 

\subsection{Implementation details}

In our study, two different architectures are adopted to investigate
the performance of spectrum anomaly detection. The architectures of
convolutional VAE and fully connected VAE are shown in Tab. \ref{tab:conv}
and Tab. \ref{tab:fully} respectively. Our network is trained with
Adam optimizer with $\alpha=0.001,\,\beta_{1}=0.9,\,\beta_{2}=0.999$.
We train the network with batch size of 32 and implement it using
Keras. It is worth noting that the VAE network is trained by normal
examples in an unsupervised approach.
\begin{table*}[tbh]
\caption{\label{tab:conv}Architecture of convolutional VAE.}

\centering{}%
\begin{tabular}{cccccc}
\hline 
Section & Layer & Type & Kernel & No. of filters & Activation function\tabularnewline
\hline 
\multirow{2}{*}{Encoder} & $1,\,2,\,3$ & Conv & $\left(2,\,2\right)$ & $2,\,4,\,8$ & LeakyReLU\tabularnewline
 & $4,\,5$ & Fully connected & $4096,\,75$ & - & LeakyReLU\tabularnewline
\hline 
\multirow{2}{*}{Decoder} & $1,\,2,\,3$ & DeConv & $\left(2,\,2\right)$ & $8,\,4,\,2$ & LeakyReLU\tabularnewline
 & $4$ & DeConv & $\left(2,\,2\right)$ & $1$ & Sigmoid\tabularnewline
\hline 
\end{tabular}
\end{table*}
\begin{table*}[tbh]
\caption{\label{tab:fully}Architecture of fully connected VAE.}

\centering{}%
\begin{tabular}{cccccc}
\hline 
Section & Layer & Type & Kernel & No. of filters & Activation function\tabularnewline
\hline 
\multirow{2}{*}{Encoder} & $1,\,2,\,3$ & Fully connected & $1024,\,256,\,64$ & - & LeakyReLU\tabularnewline
 & $4$ & Fully connected & $8$ & - & Softplus\tabularnewline
\hline 
\multirow{2}{*}{Decoder} & $1,\,2,\,3$ & Fully connected & $64,\,256,\,1024$ & - & LeakyReLU\tabularnewline
 & $4$ & Fully connected & $8$ & - & Tanh\tabularnewline
\hline 
\end{tabular}
\end{table*}

\subsection{Noise floor change\label{subsec:Noise-Floor-Change}}

\begin{figure}[tbh]
\begin{centering}
\includegraphics[width=1\linewidth]{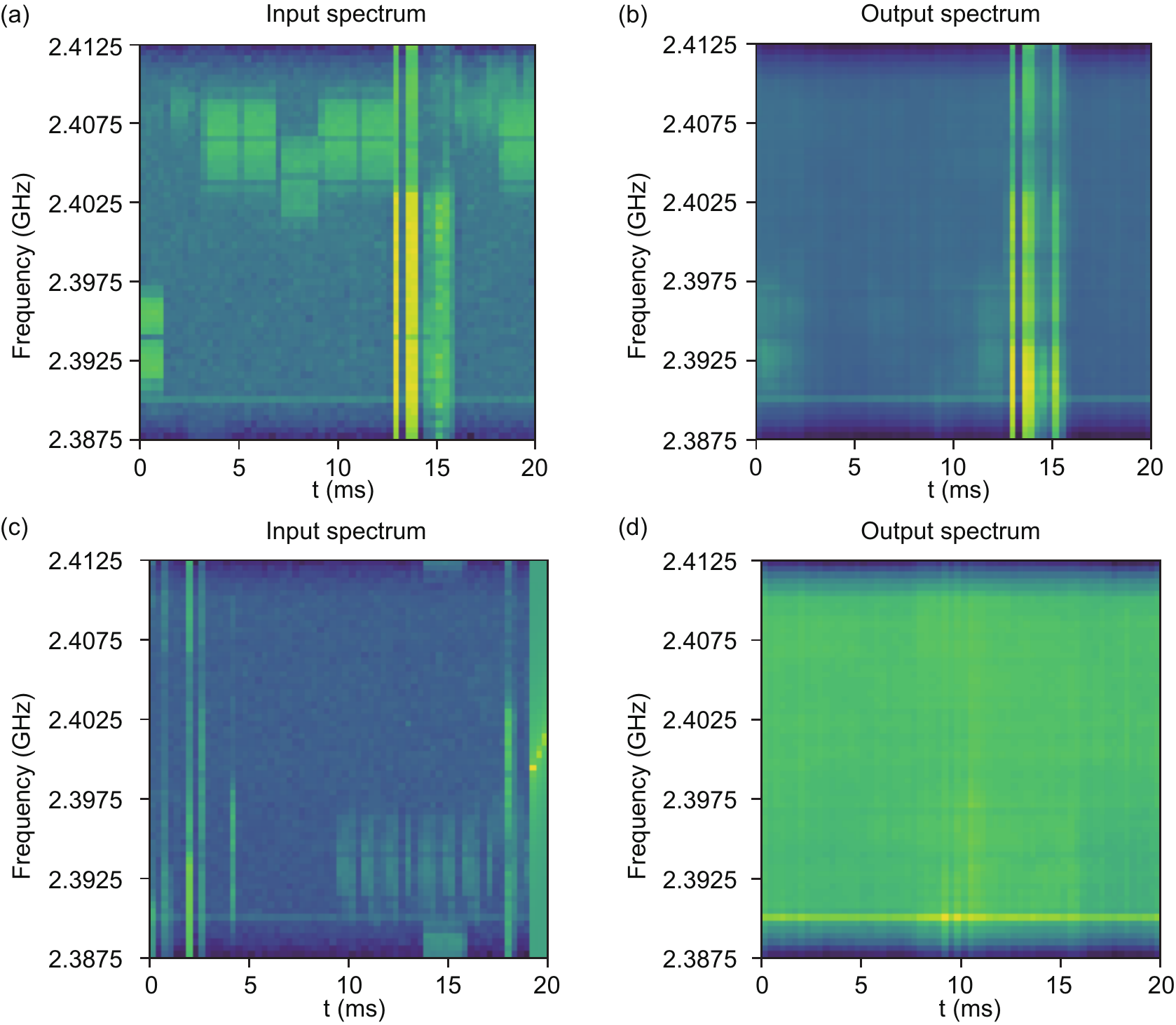}
\par\end{centering}
\caption{\label{fig:reconsturction}The input and reconstruction output of
(a-b) normal spectrum, (c-d) abnormal spectrum.}
\end{figure}

In order to illustrate the noise floor change caused by abnormal spectrum
usage, the reconstructed output of VAE of normal and abnormal samples
are compared in Fig. \ref{fig:reconsturction}. For abnormal spectrum,
the reconstruction result has a larger noise floor compared with input,
which cannot be observed in the normal case in Fig. \ref{fig:reconsturction}
(a). This is due to the fact that the decoder output approximates
the mean-average of nearby normal samples as derived in Sec. \ref{sec:Principle}.

\subsection{ROC results\label{subsec:ROC-Results}}

\begin{figure*}[tbh]
\begin{centering}
\includegraphics[width=0.84\linewidth]{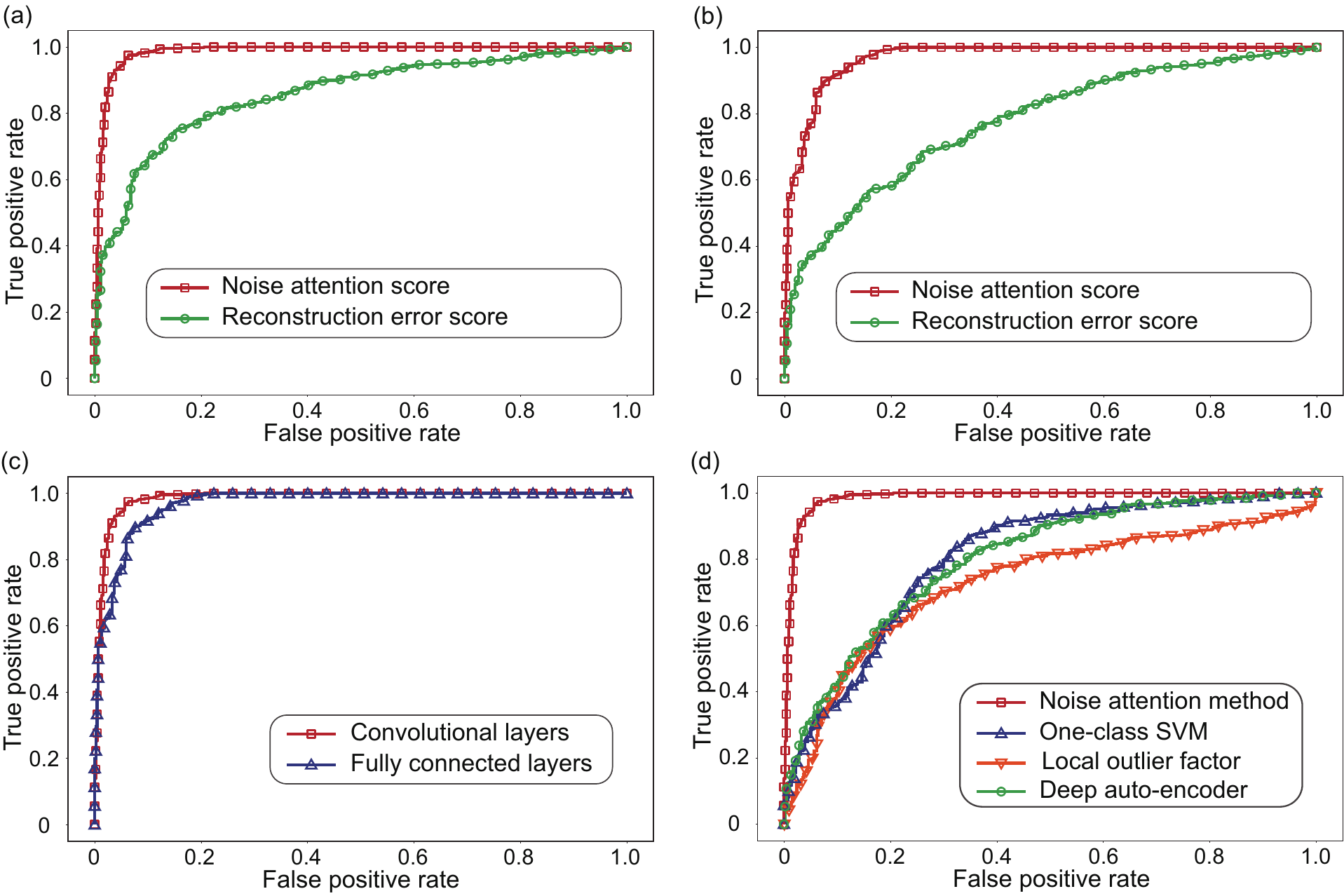}
\par\end{centering}
\caption{\label{fig:results} Performance of noise attention score on VAE with
(a) convolutional layers, (b) fully connected layers. (c) Comparison
between convolutional VAE and fully connected VAE. (d) Comparison
of noise attention based method with traditional methods.}
\end{figure*}

The performance of noise attention score is evaluated by ROC test.
ROC curve examines the variation of true positive rate (TPR) with
false positive rate (FPR). TPR equals the ratio of abnormal samples
that are detected as abnormal. FPR equals the ratio of normal samples
that are detected as abnormal, which is the false alarm rate. The
Area Under Curve (AUC) index reflects the performance of the detector.
The closer the AUC is to 1, the better the performance of the detector
is.

The performance of noise attention score and reconstruction error
score are firstly compared in convolutional VAE model. The network
architecture is shown in Tab. \ref{tab:conv}. The result is shown
in Fig. \ref{fig:results}(a). As can be seen, the proposed noised
attention score achieves higher anomaly detection performance under
same false alarm rates. AUC metrics of noise attention score and reconstruction
error score are respectively 0.986 and 0.857. The performance of noise
attention score is further examined in fully connected VAE model.
The network architecture is shown in Tab. \ref{tab:fully}. The result
is shown in Fig. \ref{fig:results}(b). The AUC metric is substantially
improved by 0.193 due to the use of noise attention score. 

The performance comparison of convolutional VAE model and fully connected
VAE model is shown in Fig. \ref{fig:results}(c). Noise attention
score is used in both methods. The AUC metric of fully connected VAE
is 0.016 lower. As can be seen, convolutional VAE model shows better
performance at low FPR region, which means that convolutional VAE
model can achieve lower false alarm rate for given detection rate. 

Finally, we compare the proposed noise attention based method with
traditional methods for spectrum anomaly detection. Three traditional
methods are considered here, which includes one-class support vector
machine (one-class SVM) \cite{Schlkopf1999Support}, local outlier
factor \cite{Breuing} and deep auto-encoder \cite{Hinton1993Autoencoders}.
Convolutional VAE mode is used in noise attention based method. Experimental
results are shown in Fig. \ref{fig:results}(d). It can be clearly
seen that the proposed method greatly outperforms the other methods.
The AUC of these methods are summarized in Tab. \ref{tab:AUC}. 

\begin{table*}[tbh]
\caption{\label{tab:AUC}The AUC of four methods.}

\centering{}%
\begin{tabular}{ccccc}
\hline 
Model & Noise attention method & One-class SVM & Local outlier factor & Deep auto-encoder\tabularnewline
\hline 
AUC & $0.986$ & $0.807$ & $0.730$ & $0.802$\tabularnewline
\hline 
\end{tabular}
\end{table*}

\section{Conclusions\label{sec:Conclusions}}

In this paper, the noise attention method is proposed for unsupervised
spectrum anomaly detection in unauthorized bands, which leverages
VAE to capture the spectrum data distribution and detects anomalies
by selectively comparing the difference between spectrogram and its
VAE reconstruction counterparts. The optimal decoder output of VAE
is theoretically derived from the perspective of the minimization
of generation error. It is indicated that the abnormal frequency usage
will elevate the noise floor in spectrogram after VAE reconstruction.
On this basis, the noise attention score is proposed for spectrum
anomaly detection, which pays more attention to the background change
in spectrograms. Experimental results show that noise attention score
significantly increases the AUC metric by 0.193, compared with the
traditional reconstruction error anomaly score. The proposed anomaly
detection method is beneficial to guide spectrum sensing system to
quickly filter out the high-value information from a large amount
of spectrum data, and automatically invest more computing and storing
resources for time-frequency window containing threatening signals.

\bibliographystyle{IEEEtran}
\bibliography{Citations}

\begin{IEEEbiography}[{\includegraphics[width=1in,height=1.25in,clip,keepaspectratio]{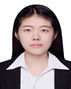}}]{Jing Xu}
 received B.S. degree in Honors College from Northwestern Polytechnical University in 2018. She is currently pursuing the M.S. degree with Qian Xuesen Laboratory of Space Technology, China Academy of Space Technology, Beijing, China. Her research interests include machine learning, spectrum sensing and data mining.
 \end{IEEEbiography}

\begin{IEEEbiography}[{\includegraphics[width=1in,height=1.25in,clip,keepaspectratio]{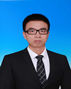}}]{Yu Tian}
 was born in 1990 in Hebei, China. He received the Ph.D. degrees in School of Electronics Engineering and Computer Science in Peking University in 2018, and received the Bachelor degrees in School of Telecommunications Engineering in XiDian University in 2013. He is currently a Co-investigator with the Qian Xuesen Laboratory of Space Technology, Beijing, China. His current research interests include Spectrum Sharing, Signal Processing, and Deep Learing.
\end{IEEEbiography}

\begin{IEEEbiography}[{\includegraphics[width=1in,height=1.25in,clip,keepaspectratio]{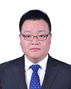}}]{Shuai Yuan}
 received the B.S. and M.S. degree from Beihang University, Beijing, China, in 2011 and 2014, respectively, both in electronic and information engineering. He is currently a research assistant in Qian Xuesen Laboratory of Space Technology, Beijing, China. His current research interests include error-control coding, spectrum sensing and processing.
\end{IEEEbiography}

\begin{IEEEbiography}[{\includegraphics[width=1in,height=1.25in,clip,keepaspectratio]{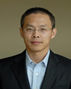}}]{Naijin Liu}
 is the deputy director of Qian Xuesen Laboratory of Space Technology, China Academy of Space Technology, Beijing, China. He received his Ph.D. degree from China University of Science and Technology. His research interests include satellite communication and space intelligent information networking. At present, he is a council member of the Chinese Society of Astronautics.
\end{IEEEbiography}

\end{document}